\documentclass[12pt]{article}
\usepackage{latexsym}
\usepackage{graphicx}
\usepackage[english]{babel}
\usepackage{amsmath}
\usepackage{setspace}
\begin{document}

\title{ A model of the late universe with viscous Zel'ldovich fluid and decaying vacuum}

\author{Rajagopalan Nair K and Titus K. Mathew, \\ Department of Physics, \\ Cochin University of Science and Technology, \\ Kochi-22, India.}

\date{}

\maketitle

\section{Introduction}
\label{intro}

The recent cosmological observations have revealed that the present universe is accelerating in expansion \cite{Perl99,Riess98} due to the presence of an exotic component called 
dark energy which constitutes more than seventy percent of the total cosmic density. Earlier to this discovery, there were indications
from weak \cite{RefregierA03} and strong\cite{TysonJA98} lensing phenomena, large scale structure \cite{AllenSW03}, galaxy rotation 
\cite{ZwickyF93,RubinVC70,RubinVC80}, on the presence of another exotic component, the dark matter, which constitutes nearly 
23$\%$ of the total density. The other known components are radiation and baryonic matter, comprising only about 1$\%$ and 5$\%$ 
respectively\cite{RefregierA03,TysonJA98,AllenSW03,ZwickyF93,RubinVC70}. Till date there is no
clear understanding regarding the nature of dark energy and dark matter. Many models have been proposed to explain the nature and evolution of these 
two hitherto unknown components. In the most successful model of the universe, the standard $\Lambda$CDM model, the dark matter is 
considered 
as some non-relativistic matter of zero pressure and dark energy is considered as cosmological constant. 
Due to the severe discrepancy between the observed and predicted values of the cosmological constant, called the cosmological constant problem, which doesn't
have any natural 
explanation in the $\Lambda$CDM model, many alternative models have been considered in the recent literature with varying dark energy densities instead of 
pure cosmological constant\cite{SolaJ13}. Regarding dark matter also, there were no general consensus at all about its nature. 

Even though the current data supports the exotic components like dark matter and dark energy, it is not really ruling out the existence of other 
exotic components. For instance, there exist speculations regarding the existence of a component called "dark radiation"\cite{Dutta11}. Another 
exotic component of interest is stiff fluid. In some early papers, 
Zel'dovich\cite{Zel'dovich62,Zel'dovich72} proposed a model in which the early stage of the universe is composed of stiff gas of baryonic particles 
with equation of state $p_z=\rho_z,$ where $p_z$ is the pressure and $\rho_z$ is the density. This equation of state implies that the velocity of 
sound in such fluids will become equal to that of light\cite{Zel'dovich62}. Later many others studied the nature and evolution of such a stiff fluid 
in an expanding universe.   
Steili et al\cite{Steili10} have shown that the self
interaction field between dark matter components will behave like
stiff fluid  and can be taken as an indication that the so called dark matter can be of stiff nature at least in the early 
stages of the universe. 
On investigating the early stages of
the universe, Barrow\cite{Barrow86} was able to establish the possible
existence of Zel'dovich fluid (or stiff fluid). 
In dissipative cosmological models also, the presence of stiff fluid in the early stage of the universe has been speculated\cite{Cruz1}.
The equation of state of the Zel'dovich fluid directly implies that the energy density of it decreases as,
$\rho_z \propto a^{-6}$ as the universe expands, where $a$ is the scale factor of the universe.  Because of this, the normal 
Zel'dovich fluid would have decoupled comparatively earlier to other components like 
 radiation, which evolves as $a^{-4}$ and  matter evolving as $a^{-3}.$
 Consequently the presence of normal Zel'dovich fluid would have a dominant effect only during the early stage of the universe.
 Inspired by these, the effect of stiff
fluid on the primordial abundances of light elements was numerically
computed by Dutta and Scherrer\cite{Dutta10}. By comparing the prediction with observational abundance of light elements, especially with the 
abundance primordial helium-4, these 
authors have obtained a constraint on the Zel'dovich fluid density as,  $\rho_z / \rho_c < 30,$ where $\rho_c$ is the critical density of the universe.
Recently a simple cosmological model consisting of stiff fluid, non-relativistic matter and cosmological constant as dark energy was studied
by Chavanis\cite{Chavanis1} and he has shown that, the universe began with a big-bang and will end in a de Sitter phase.

Even though the conventional Zel'dovich fluid would not have any effect on the late stage of the universe, the Zel'dovich fluid with bulk 
viscosity could have strong influence on the later stage. Effect of bulk viscosity, especially on the late evolution of the universe was 
studied by many\cite{Brevik1,Athira15,Jerin1}. It is worthwhile to note the analayses of the bulk viscosity in the context of early inflation in 
reference\cite{Bamba1}.
A cosmological model with bulk viscous Zel'dovich fluid as the dominant cosmic component was analysed in
reference\cite{TitusAswatiManoj14}. 
This model predicts the late acceleration of the universe when the weighted bulk viscous coefficient is in the range $4 < \displaystyle 3\zeta /H_0 <6.$
The authors have further studied the evolution of the model in the statefinder, $(r,s)$ plane, and have shown that the current state of the universe 
is arguably different from the standard $\Lambda$CDM model. 
 The statefinder parameters are the most suitable diagnostics to contrast the model with other relevant dark energy models, expecially $\Lambda$CDM. 
The evolution of the parameters 
$r$ and $s$ and their present values are used to contrast the model. The $r-s$ plane is a two dimensional plane with $r$ and $s$ as the coordinates, 
used to show the comprehensive evolution of (these parameters) 
the model. For more details on statefinder analysis see \cite{Sahni1}.
The authors also proved that the model satisfies the generalized second law of 
thermodynamics at the Hubble horizon. This work was further extended in reference\cite{TitusRajagopal16}, where the authors have used the Union 
2 supernovae data to constrain
the possible value of the viscous coefficient and obtained $\zeta/H_0 \sim 5.25$ for $H_0 \sim 70$km/s/Mpc, where $H_0$ is the Hubble parameter of the present epoch 
of the universe. However the age of
the universe was found to be around $10$ GY only. The over all behavior of this model is such that, it asymptotically evolves to a de 
Sitter one with deceleration 
parameter, $q \to -1.$
The authors\cite{TitusRajagopal16} have substantiated their results by performing a dynamical system analysis and have shown that the asymptotic  
de Sitter epoch is stable. Another important point, evident from their work 
is that the effective equation of state of the Zel'dovich fluid would start evolving from stiff nature ($p/\rho=1$), pass through a state of  
pressureless matter ($p=0$) and eventually tend towards, $p=-\rho$ corresponding to de Sitter epoch. So the inclusion of the bulk viscosity in 
the stiff matter would naturally take the fluid to manifest as pressureless matter and subsequently as a pure cosmological constant during the successive 
stages of the evolution of the universe. In spite of this reasonable evolution of the universe in the bulk viscous Zel'dovich fluid, the odd man out is 
the prediction regarding the age of the universe, which is less compared to the current observational results. In the present work we try to alleviate this 
drawback of the model by incorporating one more component, a varying vacuum energy as dark energy.
We found that with such an addition in the cosmic components, the age of the universe can be enhanced to around $12$ GY.

A scalar field model of the universe with both stiff fluid and an effective interacting vacuum were proposed 
by Cataldo et al.\cite{Cataldo1}. The authors have considered a scalar field in FLRW universe and shown that, the field 
can effectively manifest as a mixture two barotropic fluids, one with an equation of state $p=\rho,$ corresponding to the 
stiff fluid and the other with an equation of state $p=-\rho$ mimicking the vacuum. The self interaction of the field, which 
manifests as the interaction between the components, would allow the cosmological vacuum to become a dynamical quantity. We will 
further describes regarding this work in a later section.

In later  works, many have found that there exists an alternate route to the decaying vacuum, which comes from 
quantum field theory techniques based on the renormalization group approach in curved spacetimes. The effective
action in such theories inherits quantum effects from the matter sector. In general, the
renormalization group techniques in curved spacetimes leads to a dependence of vacuum
energy on the Hubble parameter H of the form, $\rho_{\Lambda}(H,\dot H) = M_{Pl}^2 \Lambda(H,\dot H)$ 
\cite{Sola3,Sola4,Shapiro1} and references therin).

In the present work we found that the addition of the decaying vacuum energy will not affect the asymptotic 
properties of the model, such that the end de Siter phase is still a stable one.
The paper is organized as follows. In section 2, we discuss the scalar field representation of a model with a mixture of 
stiff fluid and interacting scalar field, which motivates the discussions in the following sections. In section 3, we present the 
analytical solutions of the Hubble parameter, scale factor and the other cosmological 
parameters. In section 4, we obtain the model paramters by constraining the model with supernovae observational data and then compute the evolution of the 
different cosmological parameters. We also evaluate the age of the universe in this section. Section 5 is devoted to the study of the asymptotic properties 
of this model followed by the conclusions in section 6.

\section{Scalar field approach to stiff fluid and interacting vacuum}

Let us consider the evolution of a self interacting scalar field, $\phi.$ which is minimally coupled to gravity in flat 
isotropic and homogeneous universe. In describing the evolution of the scalar field we mainly follow the reference 
\cite{Cataldo1}, so for more details see that reference. The evolution of the scalar field is governed by 
the equations
\begin{equation}\label{eqn:phi1}
 3H^2 = 8\pi G \rho_{\phi}
\end{equation}
and 
\begin{equation}\label{eqn:phi2}
 \dot \phi \ddot \phi + 3 H {\dot \phi}^2 = \dot \phi \frac{dV(\phi)}{d\phi}
\end{equation}
where $H$ is the Hubble parameter, $V(\phi)$ is the potential, dot represents a derivative with respect to cosmic time and 
equation(\ref{eqn:phi2}) represents the dynamical evolution of the field. The equation of state of the field is,
\begin{equation}
 \omega_{\phi}=\frac{p_{\phi}}{\rho_{\phi}},
\end{equation}
in which the pressure and density of the scalar field are given by,
\begin{equation}
 \rho_{\phi}=\frac{{\dot \phi}^2}{2} + V(\phi), \, \, \, \, \, \, p_{\phi}=\frac{{\dot \phi}^2}{2} - V(\phi)
\end{equation}

The self interacting scalar field can be effectively treated as the mixture of two interacting fluids, with densities 
$\rho_1$ and $\rho_2$ which are having the equation of state $\omega_1$ and $\omega_2$ respectively. Then the effective 
pressure is,
\begin{equation}
 P_{eff}=\omega_{eff} \rho_{eff} = \frac{\omega_1 \rho_1 +\omega_2 \rho_2}{\rho_1+\rho_2} \rho_{eff},
\end{equation}
where $\rho_{eff}=\rho_1+\rho_2.$
Hence it is now possible to have two fluids, one with equation of state, $\omega_1=1,$ corresponding to stiff fluid and 
the other with equation of state, $\omega_2=-1$ corresponding to vacuum energy, if one identifies the corresponding 
densities and pressures as,
\begin{equation}\label{eqn:rhop1}
 \rho_1=\frac{{\dot \phi}^2}{2}, \, \, \, \, \, p_1=\frac{{\dot \phi}^2}{2}
\end{equation}
and 
\begin{equation}\label{eqn:rhop2}
 \rho_2=V(\phi), \, \, \, \, \, p_2=-V(\phi).
\end{equation}
One can either take these components as isolated from each other, so that each one of them satisfying separate 
conservation laws or can be taken as interacting components following the conservation equations,
\begin{equation}\label{eqn:con123}
 {\dot \rho}_1 + 3 H \left(\rho_1+p_1\right)= Q, \, \, {\dot \rho}_2 + 3 H \left(\rho_2+p_2\right)= -Q,
\end{equation}
for $Q>0$ the energy flows for $\rho_2$ to $\rho_1$ and for $Q<0$ the energy flow in the reverse direction. From equations 
(\ref{eqn:rhop1}),(\ref{eqn:rhop2}) and (\ref{eqn:con123}), it follows that,
\begin{equation}
 {\dot \phi} \ddot \phi + 3 H {\dot \phi}^2 = Q(t), \, \, \, \, \dot \phi \frac{dV(\phi)}{d\phi}=-Q.
\end{equation}
These equations are the equivalent evolution equation of the scalar field. Hence it is possible to consider a scalar field 
as a mixture of a stiff fluid interacting with an effective vacuum energy.

In reference \cite{Cataldo1}, the authors have described the evolution of such a universe. Our aim is not in line with that. 
Instead we consider a phenomenological form for the decaying vacuum along with the bulk viscous stiff fluid (Zel'dovich fluid) and 
analyze both the background evolution of the universe, particularly in finding the age and also the asymptotic behavior.

\section{Background evolution with viscous Zel'dovich fluid and decaying vacuum $\Lambda(t)$}

We consider a flat FLRW universe with standard metric
\begin{equation}\label{eqn:metric}
ds^2=-c^2dt^2+a^{2}(t)\left(dr^2+r^2 d\theta^2+r^2sin^2\theta \,
d\phi^2 \, \right)
\end{equation}
where $t$ is the cosmic time, $(r,\theta,\phi)$ are the co-moving
coordinates and $a(t)$ is the scale factor. The cosmic components are the
bulk viscous Zel'dovich and a varying cosmological constant. A non-viscous ideal Zel'dovich fluid 
obeys the equation of state $p_z=\rho_z$\cite{Zel'dovich62}, which will change once the fluid is assumed to be viscous. 
Bulk viscosity can be arised due to the deviations of the system from local thermodynamic equilibrium\cite{Zimdahl1}. It arises as an 
effective pressure to restore the system back to its equilibrium, whenever the cosmic fluid expands or contracts too fast.  
In incorporating the viscosity into the analysis we adopt 
the Eckart's formulation\cite{Eckart40}.  Landau and Lifshitz also discussed a formulation\cite{Landau58} equivalent to Eckart's.
There had been works showing that the thermodynamical equillibria in Eckart's theory as unstable\cite{Hiscock85}  and signals could propagate through the 
fluid at superluminal velocities\cite{Israel76}. But these difficulties can be overcome by taking account of the higher order terms. Such a more general 
formalism was developed by Israel et al \cite{Israel791,Israel792}, to which Eckart's theory would appear as a first order limit. However owing 
to the simplicity many authors favor Eckart's formalism for a first step analysis. 
Later Hiscock and Salmonson \cite{Hiscock91} had shown that Eckart's formalism can be reasonably applied to  FLRW universe with late acceleration. 

 For a first order deviation from thermodynamical equilibrium  
the energy momentum tensor can be written as,
\begin{equation}
T^{\mu\nu}=\rho u^{\mu} u^{\nu} + \left(p+\Pi\right)h^{\mu\nu}
\end{equation}
where 
 $\rho$ is the density of the fluid component, $u^{\mu}$ is the four velocity of an observer in Hubble flow,
 $\left(p+\Pi\right)$ is the effective pressure and  $h^{\mu \nu}=u^{\mu} u^{\nu}+g^{\mu \nu},$ with $g^{\mu\nu}$ as the metric coefficients.
 The above mentioned approach results in the equation for 
 effective pressure as $p^{\prime}=p+\Pi$ for the bulk viscous Zel'dovich fluid and it can be expressed as,
\begin{equation}\label{eqn:pzdash}
p^{\prime}_z=p_z-3\zeta H
\end{equation}
where we have $p=p_z$ the normal pressure, $\Pi=-3\zeta H,$ the viscous pressure and $H$ is the Hubble parameter. 
 In cosmology, bulk viscosity arises as an effective
pressure to restore the system back to its thermal equilibrium, which is broken when the cosmological fluid expands
(or contracts) too fast. This bulk viscosity pressure generated
ceases as soon as the fluid reaches the thermal equilibrium\cite{Wilson1}.

The second cosmic component in the present model is the time varying cosmological parameter given by\cite{Bessada1},
\begin{equation}
\Lambda(t)=3\alpha H^2
\end{equation}
where $\alpha$ is a free parameter, value of which would be less than one.  
 Earlier 
introduction of this kind of decaying vacuum was consdiered by Carvalho and Lima\cite{Carvalho1}, where the authors restricted to $\alpha \leq 1/2.$ 
A higher 
value of $\alpha$ resulted into incompatible age for the universe as claimed by many authors like \cite{Bessada1}, so the values of $\alpha$ is usually 
restricted to below one.
Since this is effectively a form of time varying 
vacuum energy,
its equation of state is taken as, $ \displaystyle \omega_{\Lambda}=(p_{\Lambda}/\rho_{\Lambda})=-1.$ Basically the $\Lambda(t)$ 
models have been originated from curved space quantum field theories\cite{SolaJ11}. Often there appears a constant additive term along 
with the time varying part in the equation for $\Lambda(t),$ which, as argued by many
\cite{SolaJ13,Paxy1} facilitate the transition from the decelerating to an accelerating epoch of the expanding universe. But in the present model such
an additive constant is not needed due to the presence of viscosity in the Zel'dovich fluid component, which will otherwise guarantee such a transition 
from an early deceleration to a later accelerating phase of expansion.

The Friedmann metric along with the standard Einstein's field equation will give the Friedmann equation for a flat universe as,
\begin{equation}\label{eqn:FrRaEq1a}
3H^2(t)=\rho_z + \rho_{\Lambda}
\end{equation}
where $\rho_{\Lambda}=\Lambda(t)$ (in standard units, $8\pi G=1, \, c=1$), is the time varying cosmological parameter, 
equivalent to the standard dark energy density. These components together satisfy the conservation law (in the absence of 
any source, i.e. $Q=0$),
\begin{equation}
\dot{ \rho}_{\Lambda} + \dot{\rho}_z+3 H \left(\rho_z + p'_z \right)=0.
\end{equation}
where a dot represents a derivative with respect to cosmic time.
Combining the Friedmann equation and the conservation equation,  leads to
\begin{equation}\label{eqn:dotRho2}
\dot{\rho}_z+\dot{\rho}_{\Lambda}= - 6 H\left(\rho_z-\frac{3}{2}\zeta H\right).
\end{equation}
But from Friedmann equation, $\dot{\rho}_z+\dot{\rho}_{\Lambda} = 6 H \dot H.$ On substituting this, the above equation becomes,
\begin{equation}\label{eqn:dotH}
\dot{H}=- \left(\rho_z-\frac{3}{2}\zeta H\right).
\end{equation}
Again from Friedmann equation one can substitute for $\rho_z$ as,
\begin{equation}\label{eqn:rhoz2}
\rho_z=3\left(1-\alpha \right)H^2.
\end{equation}	
From equations (\ref{eqn:dotH}) and (\ref{eqn:rhoz2}),
\begin{equation}\label{eqn:diffEqH}
\dot{H}+3H\left((1-\alpha)H-\frac{\zeta}{2}\right)=0.
\end{equation}
Solving equation (\ref{eqn:diffEqH}) we obtain the Hubble parameter as,
\begin{equation}\label{eqn:Hubble0}
H(t)=\eta \left[1+\coth\left(3(1-\alpha)\eta(t-t_0)+\phi \right) \right]
\end{equation}
where $\displaystyle \eta=\frac{\zeta}{4(1-\alpha)}$, $\displaystyle \phi=\coth^{-1}\left(\frac{H_0}{\eta}-1\right)$ and $H_0$ is the current 
value of Hubble parameter. Integrating the above equation, we obtain the equation for the scale factor as
\begin{equation}\label{eqn:aoft0}
a(t)= e^{\eta(t-t_0)}\left(\frac{\sinh[3\eta(1-\alpha)(t-t_0)+\phi]}{\sinh(\phi)}\right)^{\frac{1}{3(1-\alpha)}}.
\end{equation}
Using this equation, the Hubble parameter in equation (\ref{eqn:Hubble0}) can be recast as,
\begin{equation}\label{eqn:Hubble1}
H=\frac{\zeta}{2(1-\alpha)}+\left(H_0-\frac{\zeta}{2(1-\alpha)}\right)a^{-3(1-\alpha)}.
\end{equation}
In the asymptotic limit $a(t) \to \infty $ the Hubble parameter becomes a constant, 
$\displaystyle H \to \frac{\zeta}{2(1-\alpha)}$ which corresponds to the de Sitter phase with exponential increase in the scale factor,  
while in the limit $a(t) \to 0,$ the Hubble parameter 
evolves as, $H \sim a^{-3(1-\alpha)},$ which points to an earlier decelerated epoch dominated with Zel'dovich fluid with density 
$\rho_z \sim H^2 \sim a^{-6(1-\alpha)}.$ Thus the existence of the transition from an early decelerated to a late 
accelerated epoch is guaranteed. In these limits the scale factor will evolve as follows. 
As $t \to \infty$ the scale factor will evolve as, 
$a(t) \to e^{2\eta(t-t_0)},$ this exponential increase corresponds to the de Sitter epoch, while in the early stage of the evolution, corresponding to 
$3\eta(1-\alpha)(t-t_0)<1$ the above form of 
$a(t)$ almost implies that, $a(t) \sim \left((1+3\eta(1-\alpha)(t-t_0)\right)^{1/3(1-\alpha)},$ representing a decelerating epoch.
 What is important here is that the transition occurs  
  without the aid of an additive constant in the $\Lambda(t).$ In non-viscous models like
entropic dark energy\cite{Basilokos1} or Ricci dark energy\cite{Paxy1}, the presence of a bare constant cosmological term is essential for having a 
transition from the early decelerated epoch to the late accelerated epoch. 

The evolution of the cosmological parameters, like Hubble parameter, scale factor etc are depending upon the numerical values of the model 
parameters $\alpha$ and  
viscous coefficient $\zeta.$ However it is clear from the expression of scale factor in equation(\ref{eqn:aoft0}) that for a constant $\alpha$ 
the beginning of the universe corresponding to 
 $a=0$ would have occurred earlier into the past of the universe as $\zeta$ assumes higher values. 
For constant $\zeta$ and increasing $\alpha,$ the situation will be the same too. In both the cases the age of the universe increases 
compared to a model with only Zel'dovich fluid as the cosmic component. However, only with an extraction 
of these parameters, a final conclusion regarding the age of the universe can be made.

  We have considered a flat universe ($k=0$), since observations
 strongly indicate that our universe is flat \cite{Komatsu1}. 
 The inflationary models theoretically propose a very small value for curvature around $\Omega_{k0} \sim 10^{-5}$ while observations 
 favor a value of the order of $10^{-2}.$ 
 Basically 
 for non-flat universe, the Friedmann equation becomes,
 \begin{equation}\label{eqn:frw11}
  3H^2 = \rho_z + \rho_{\Lambda}+\rho_k.
 \end{equation}
 where $\rho_k=-ka^{-2}.$
Since the 
interaction is only between Zel'dovizh fluid and the vacuum, the conservation law is,
\begin{equation}\label{eqn:cons11}
 \dot\rho_z + \dot{\rho}_{\Lambda} + 3 H \left(\rho_z+p^{\prime}_z\right)=0.
\end{equation}
Using equations (\ref{eqn:pzdash}) and (\ref{eqn:rhoz2}) and through simple algebra we can rewrite the above equation as,
\begin{equation}
 \dot H +3H\left( (1-\alpha)H-\frac{\zeta}{2}\right)=\dot{\rho}_k.
\end{equation}
In our original work we took, $\rho_k=0$ consequently the rhs of the above equation is zero. But for non-flat universe the contribution due to the 
rhs term, $\dot{\rho}_k=-2H\rho_k$ is extremly small especially in the late stage, first of all due to the decreasing nature of $H$ and secondly 
due to the extremely low magnitude of $\rho_k.$  Hence the solution of the above equation (our model) would be almost close to the solution of the 
corresponding homogeneous equation with zero curvature.

\section{Extraction of model parameters and evolution of cosmic parameters }
The best fit values for $\zeta,\alpha$ and $H_0$  are estimated using type Ia supernova observational data. Union data 
containing 307 data points \cite{Kovalsky08} 
in the red shift range $0.01<z<1.55$ has been used here. For $i^{th}$  supernova  at a red shift $z,$ having 
an apparent magnitude $m$ and absolute magnitude
$M,$ the distance modulus is,
\begin{figure}[h]
\centering
\includegraphics[scale=0.5]{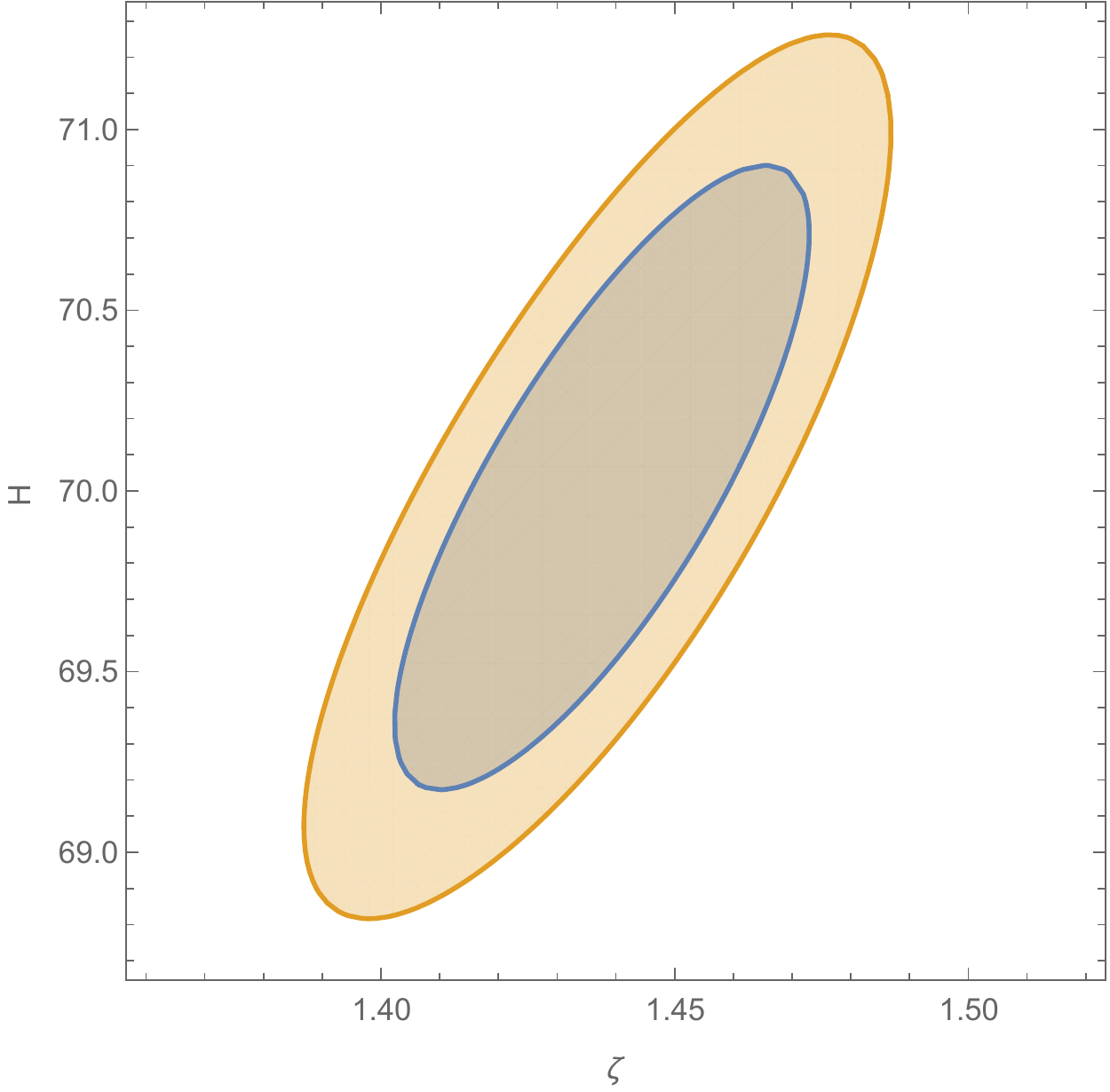}
\caption{Contour plot for the parameters $H$ and $\bar\zeta$ for fixed $\alpha=0.14.$}
\label{fig:cont1}
\end{figure}
\begin{equation}\label{eqn:distancemodulus}
\mu_i (z)=(m-M) = 5\log_{10} d_L(z)+25,
\end{equation}
where $d_L(z)$ is its luminosity distance, which is depending on the model parameters also and is given by 
\begin{equation}\label{eqn:dLz}
d_L(z,\bar{\zeta},\alpha)=\frac{c(1+z)}{H_0} \int_0^z \frac{dz^{'}}{h(z^{'},\bar{\zeta},\alpha)},
\end{equation}
where $\displaystyle h(z^{'},\bar\zeta,\alpha)=\frac{H(z^{'},\bar\zeta,\alpha)}{H_0}.$
Here we have redefined the viscosity coefficient by weighting it with the present Hubble parameter as, $\displaystyle \bar{\zeta}=\frac{\zeta}{H_0}.$
Equation for $h$ in terms of $z,$ the cosmological red shift is obtained by substituting for scale factor using $a=\frac{1}{1+z}$ in equation(\ref{eqn:Hubble1})
as
\begin{equation}\label{eqn:h}
h(z^{'},\bar\zeta,\alpha)=\frac{\bar{\zeta}}{2(1-\alpha)}+\left(1-\frac{\bar{\zeta}}{2(1-\alpha)}\right)(1+z)^{3(1-\alpha)}.
\end{equation}

The theoretical distance moduli for various red shifts are obtained using equation (\ref{eqn:dLz}) and are
compared with the corresponding observational data. The statistical $\chi^2$ function for comparing the theoretical and observational values of 
the distance moduli is defined as,
\begin{equation}
\chi^2=\sum_{i=1}^n {\left[\mu_{ith}-\mu_{iob} \right]^2 \over \sigma_i^2}
\end{equation}
where $\mu_{ith}$ is the theoretical value of the distance modulus of the $i^{th}$ supernova for a given redshift and $\mu_{iob}$ is its observed distance 
modulus corresponding to the same redshift.  $\sigma_i$ is the variance of the measurement of $i^{th}$ supernova and $n=307$ is the number of data points. 
$\mu_{ith}$ being dependent on $(\alpha,\bar{\zeta},H_0),$ the best estimates of the parameters $(\bar\zeta,\alpha,H_0)$ are obtained by
minimizing the $\chi^2$ function. The minimum of the $\chi^2$ indicates the 
goodness-of-fit of the model apart from giving the
best estimates of the model parameters.

For obtaining the $\chi^2$ function we also used 
Background (CMB) data from the WMAP 7-yr observation and the Baryon Acoustic Oscillation
(BAO) data from Sloan Digital Sky Survey(SDSS). The BAO signal has been directly
detected by SDSS survey at a scale $\sim$100MPc. The BAO peak parameter value was first proposed by
Eisenstein, D. J. et al\cite{eisenstein} and is defined as
\begin{equation}\label{equ:17}
\mathcal{A}=\frac{\sqrt{\Omega_{m}}}{h(z_{1})^{\frac{1}{3}}}\left(\frac{1}{z_{1}}\int_{0}^{z_{1}}\frac{dz}{h(z)}\right)^{\frac{2}{3}}
\end{equation}
Here  h(z) is the Hubble parameter, $z_{1} = 0.35$ is the red shift of
the SDSS sample. Using SDSS data from luminous red
galaxies survey the value of the parameter $\mathcal{A}$(for flat
universe) is given by $\mathcal{A} = 0.469 \pm 0.017$\cite{eisenstein}. The
$\chi^{2}$ function for the BAO measurement takes the form
\begin{equation}\label{equ:18}
\chi^{2}_{BAO}=\frac{(\mathcal{A}-0.469)^{2}}{(0.017)^{2}}
\end{equation}
The CMB shift parameter is the first peak of CMB power
spectrum\cite{bond} which can be written as
\begin{equation}\label{equ:19}
\mathcal{R}=\sqrt{\Omega_{m}}\int_{0}^{z_{2}}\frac{dz}{h(z)}
\end{equation}
Here $z_{2}$ is the red shift at the last scattering surface. From
the WMAP 7-year data, $z_{2}=1091.3$. At this red shift
$z_{2}$, the value of shift parameter would be $\mathcal{R}=1.725\pm
0.018$\cite{Komatsu1}. The $\chi^{2}$ function for the CMB measurement can be
written as
\begin{equation}\label{equ:20}
\chi^{2}_{CMB}=\frac{(\mathcal{R}-1.725)^{2}}{(0.018)^{2}}
\end{equation}
Considering three cosmological data sets together, i.e. (SNe+BAO+CMB),  the total $\chi^{2}$ function is
then given by
\begin{equation}\label{equ:21}
\chi^{2}_{total}=\chi^{2}_{SNe}+\chi^{2}_{BAO}+\chi^{2}_{CMB}
\end{equation}

 By minimising the $\chi^2,$ we found parameter values as, $\alpha=0.14, H_0=70.3\textrm{km/s/Mpc}, \, \textrm{and} \, 
 \bar\zeta=1.446$ for $\chi^2$ per degrees 
of freedom, $\chi^2_{dof}= \frac{\chi^2_{min}}{n-m}=1.016,$ where $n$ is the number of data points and $m=3$ the number of free parameters. 
We have constructed the confidence interval plane for the parameters $H$ and $\bar\zeta$ keeping $\alpha=0.14,$ its best estimated value. The confidence 
intervals corresponding to 68.4$\%$ and 95.4$\%$ show fairly good behavior and are given in figure.\ref{fig:cont1}. The best 
fit values for the parameters $H_0$ and $\bar \zeta$ with corrections for a confidence of 64.8$\%$ are 
$H_0=70.03_{-0.46}^{+0.54}$ and $\bar\zeta=1.446_{-0.023}^{+0.018}. $ For 95.4$\%$ probability the corrected parameter values are 
$H_0=70.3^{+0.565}_{-0.47}$ and $\bar\zeta=1,446^{+0.095}_{0.032}$ for $\alpha=0.14.$
It may be noted that in reference 
\cite{Bessada1}, the authors have extracted an upper limit for the parameter $\alpha$ by constraining a model with a decaying 
vacuum, $\Lambda=\Lambda_0+3\alpha H^2,$ 
as $\alpha \leq 0.15.$ 

In discussing the evolution of different cosmological parameters, it is better to start with the equation of state parameter. As it was shown in some 
of the earlier works\cite{TitusAswatiManoj14}, the equation of state of the viscous Zel'dovich fluid has natural evolution from its extreme 
stiff nature (corresponds $\omega=1$) 
to the de Sitter type behavior through radiation (corresponds to $\omega=1/3$) like and matter like (corresponds to $\omega=0$) natures. 
First we will consider the net equation of state, comprising both the Zel'dovich fluid and the decaying vacuum, which can be obtained by 
the standard procedure as,
\begin{equation}\label{eqn:Omega1}
\omega(z)=-1-\frac{1}{3}\frac{d}{dx}(\ln {h^2})
\end{equation}
From the equations (\ref{eqn:h})and (\ref{eqn:Omega1}), the equation of state can be expressed as,
\begin{equation}\label{eqn:Omega2}
\omega(z)=-1+\left(2(1-\alpha)-\bar\zeta\right)\frac{1}{h}(1+z)^{3(1-\alpha)}.
\end{equation}
For  $\alpha=0$, and $\bar\zeta=0$ equation of state tends to $\omega(z) \to 1$ for very large redshift, which corresponds to the early epoch dominated with 
Zel'dovich fluid with negligible viscosity. 
\begin{figure}[h]
	\centering
	\includegraphics[scale=1]{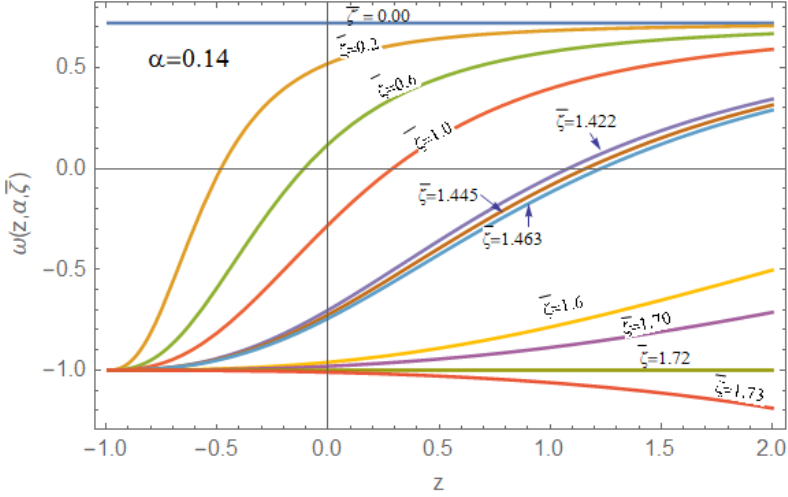}
	\caption{Evolution of $\omega$ with time for constant $\alpha=0.14$ and varied values of the viscous coefficient, $\bar\zeta.$}
	\label{fig:OAfx}
\end{figure}
For the chosen value of the parameter, $\alpha=0.14$ a transition into the late accelerating phase would occur for a range of values of the 
viscous coefficient, $1.72>\bar\zeta >0.$   For the best fit values of the parameters, $(\alpha,\bar\zeta,H_0)=(0.14,1.445,70.03)$ 
 the evolution of the equation of state parameter is as expected (see figure \ref{fig:OAfx}), that is, 
 starting with a value corresponding to stiff fluid and gradually approaching the value corresponding to de Sitter epoch. 
The current value of $\omega$ corresponding to the best fit values of the parameters is found to be  $\omega \sim -0.72,$ 
very much close to the range of values of the equation of state from WMAP data\cite{Komatsu1}. 

The equation of state of the viscous Zel'dovich fluid can be obtained using a similar approach as,
\begin{equation}
 \omega_z = -1 - \frac{1}{3} \frac{d\Omega_z}{dx}=-1 - \frac{(1-\alpha)}{3} \frac{d\ln h^2}{dx}.
\end{equation}
The difference in $\omega_z$ compared to $\omega(z)$ is that the second term on the right hand side of the above equation 
contains an extra term, $(1-\alpha),$  so that the final expression for $\omega_z$ becomes,
\begin{equation}
 \omega_z=-1+(1-\alpha) \left(2(1-\alpha)-\bar\zeta\right)\frac{1}{h}(1+z)^{3(1-\alpha)}.
\end{equation}
The general evolution of $\omega_z$ is hence similar to that of $\omega(z)$ except in the particular numerical values corresponding to 
different epochs. But both 
will approach the de Sitter value as $a \to \infty.$ 

The deceleration parameter $q(z)$ is a measure of the acceleration and can be obtained from the basic equation,
\begin{equation}\label{eqn:q0}
q=-1-\frac{\dot H}{H^2}
\end{equation}
\begin{figure}[h]
\centering
\includegraphics[scale=0.7]{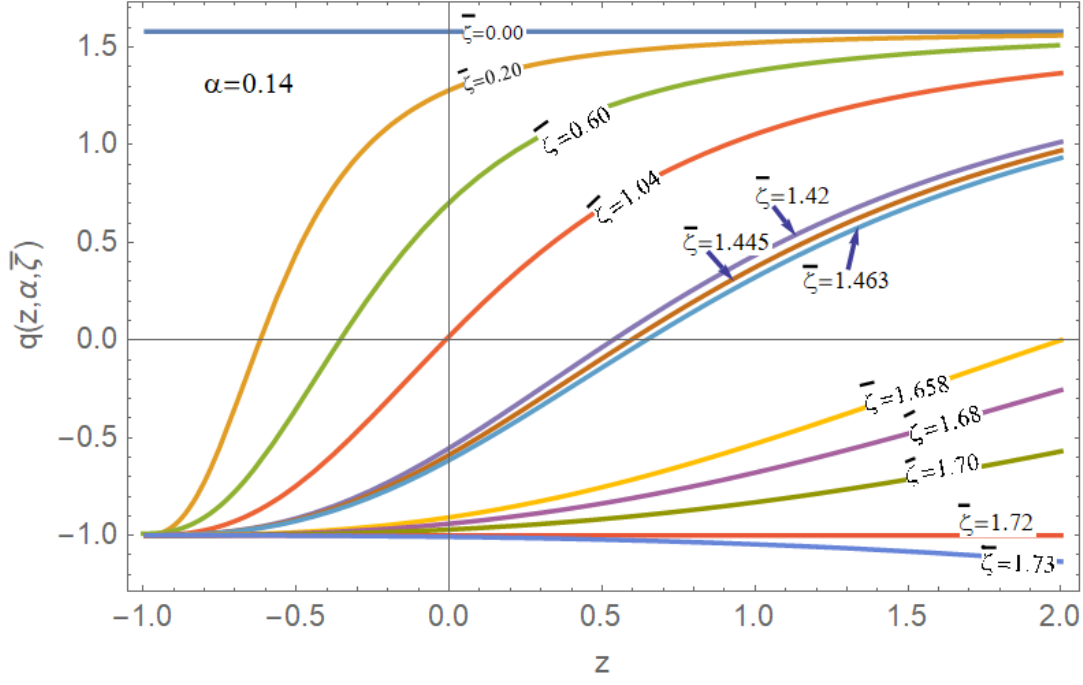}
\caption{Evolution of deceleration parameter with $\alpha=0.14$ and varying $\bar\zeta.$ }
\end{figure}
is also evaluated. Using equation(\ref{eqn:Hubble1}),  deceleration parameter takes the form
\begin{equation}
q=-1+\left(\frac{3(1-\alpha)}{1+(\frac{1}{\frac{2(1-\alpha)}{\bar\zeta}-1})(1+z)^{-3(1-\alpha)}}\right)
\end{equation}
When both the model parameters $\alpha$ and $\bar\zeta$ are equal to zero, the cosmic component becomes pure Zel'dovich fluid and $q=2.$
For the best estimated values of the model parameters, the transition is found to occur at a redshift, $z\approx 0.61$ 
which is again close to the observed value. Irrespective of the values of $\bar\zeta$ the deceleration parameter asymptotically 
approaches the value, $q=-1.$ 

\begin{figure}[h]
	\centering
	\includegraphics[scale=0.5]{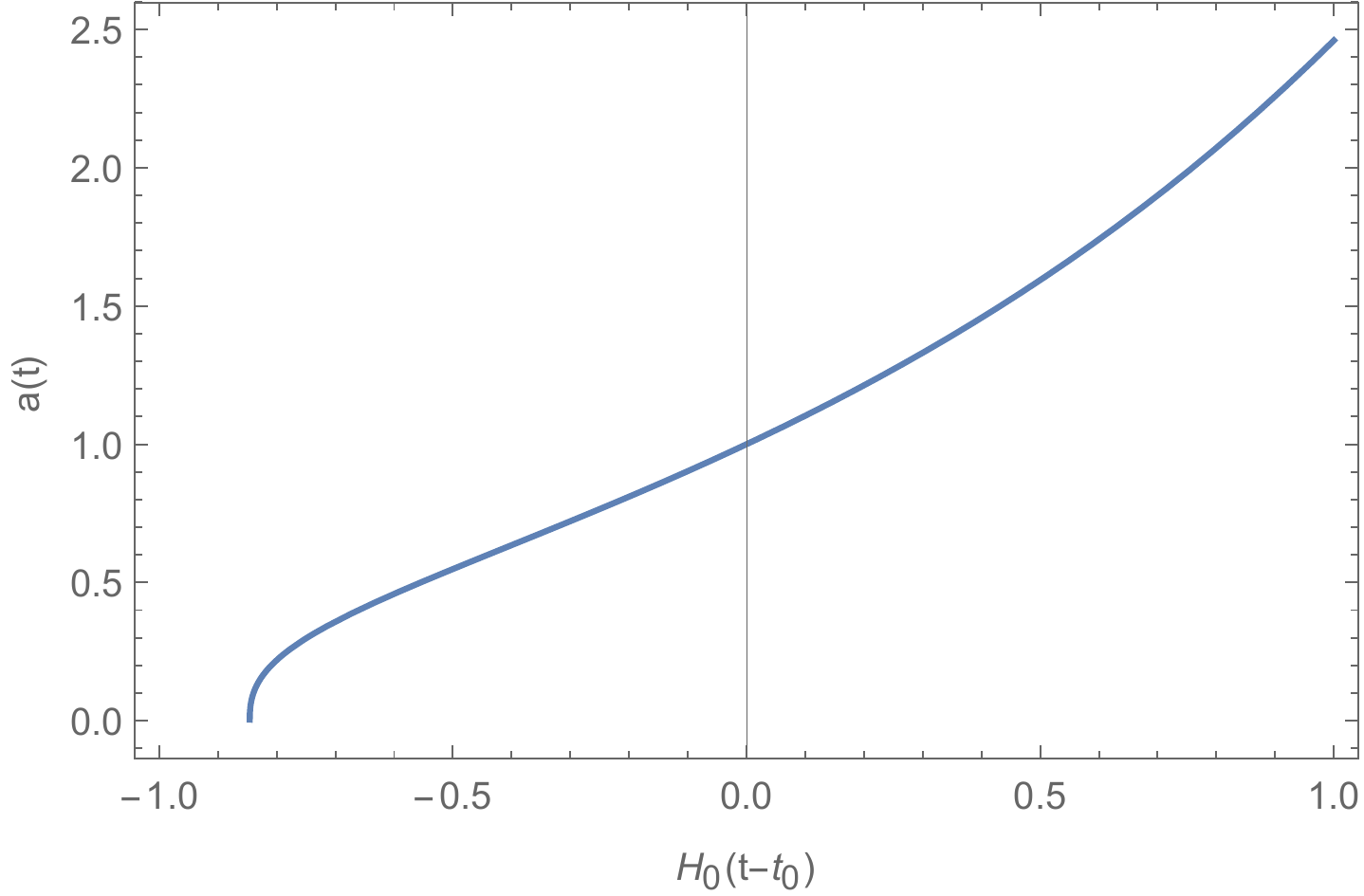}
	\caption{Evolution of scale factor with time. The profile corresponds to best fit values, $\alpha=0.14$, 
	$\bar\zeta=1.445$ and $H_0=70.03$ $km^{-1}s^{-1}Mpc^{-1}.$}
	\label{fig:a}
\end{figure}

\begin{figure}[h]
\centering
\includegraphics[scale=0.8]{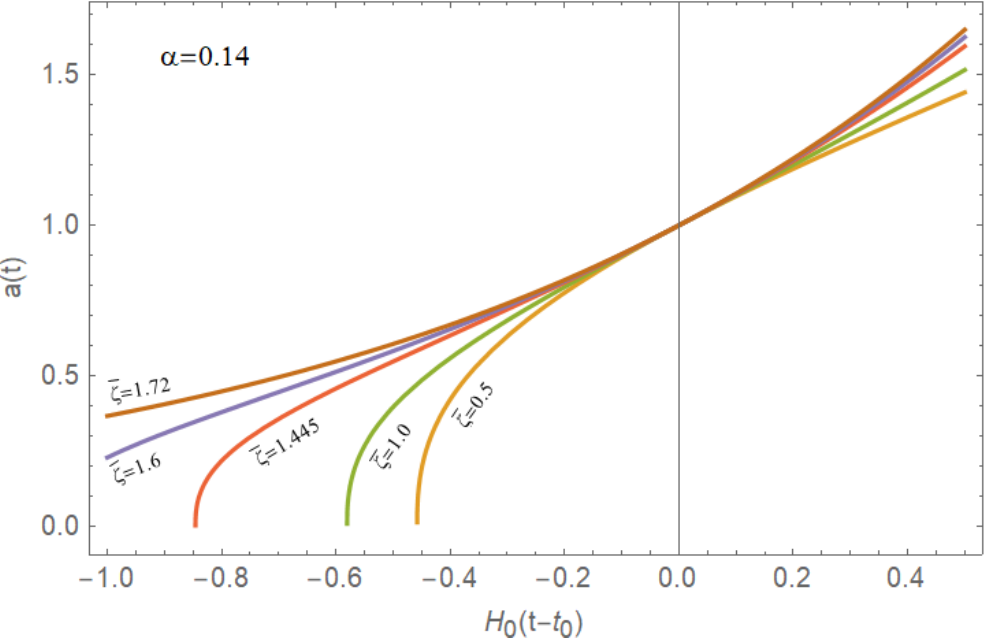}
\caption{The spectrum of curves on evolution of $a(t)$ with $\alpha=0.14$ and varying $\bar\zeta.$  
There is big-bang for $\bar\zeta<1.72.$}
\label{fig:aAlphaFx}
\end{figure}

Finally we will discuss about the evolution of the scale factor. The age of the universe can be directly obtained for the evolution of the scale factor.
The evolution of the scale factor is given in equation(\ref{eqn:aoft0}). We have already shown its asymptotic behavior in a previous section, that 
in the early epoch it evolves as in the decelerated phase and in the extreme future it evolves as in de Sitter epoch. In general,
the form of $a(t)$ indicates 
the presence of big-bang as $(t-t_0) \to -\infty.$
The evolution of it as shown in Figure (\ref{fig:a}) for the best fit values  of the model parameters indicate it. But for higher values of $\bar\zeta$ it 
is found that the big-bang occur at earlier times as evident from figure (\ref{fig:aAlphaFx}).
It is also clear that, for extremely high values of the viscous parameter, the scale factor would have no-zero values in the beginning 
indicating the absence of big-bang. It is found that, there is no big-bang in the model for $\bar\zeta>1.72.$ 
This 
means that age of the universe defined only for $\bar\zeta<1.72.$  

The age of the universe in the present model can be obtained by equating the scale factor to zero, which leads to,
\begin{equation}
 Sh \left[3\eta(1-\alpha)(t_B-t_0)+\phi\right]=0.
\end{equation}
where we took, $t=t_B,$ as the big bang time.
This leads to the equation of the age of the universe as,
\begin{equation}
 t_0-t_B=\frac{\phi}{3\eta(1-\alpha)}.
\end{equation}
On substituting the expressions for $\phi$ and $\eta,$ the above equation can be simplified into,
\begin{equation}
 t_0-t_B=\frac{2}{3\bar\zeta} \ln \left(\frac{1}{1-2\bar\eta}\right) H_0^{-1},
\end{equation}
where $\bar\eta=\eta/H_0.$ For the best estimated values of the model parameters, it is found that, the above equation gives an age of the universe 
in the range,
 $11.39-12.18$ GY. First of all, this age is higher than the age predicted by the model in which  
 Zel'dovich fluid is the dominant component\cite{TitusRajagopal16}. Secondly it is near to the age of the universe obtained from the 
 data on oldest globular clusters\cite{Pont1,CarrettaAge00, SalarisM97,SukyongYi01}. In this sense the model seems to solve the problem of age which 
 existed in the model with Zel'dovich fluid as the only cosmic component. So it can be concluded that, the inclusion of a varying dark energy component 
 along the bulk viscous Zel'dovich fluid is essential for the consistency with the age determination of the present universe.

 The analysis so far reveals that the inclusion of the additional component, the decaying vacuum to the viscous Zel'dovich fluid, the model gives 
a reasonable back ground evolution of the universe. Apart from this, the age of the resulting universe will be high compared to the 
model with Zel'dovich fluid 
alone as the cosmic component. The viscous Zel'dovich fluid component evolves in such a way that, during the very early period the matter 
component is a stiff 
fluid, compatible with many theoretical speculations\cite{Dutta10}. But as the universe evolves, the equation of state is smoothly evolving 
towards that of pressureless fluid, 
corresponding to the non-relativistic matter. Hence the model supporting the speculation that in the early period the matter 
would have existed as a stiff fluid. In 
the late stage, the evolution is compatible with the standard $\Lambda$CDM model, in predicting the observational parameters, 
including the age of the universe. In next section
we do a dynamical system analysis of the model, which may throw more light on the viability of the present model.

\section{Dynamical system analysis}
 Dynamical system analysis is an effective method to extract the useful information about the stability of the asymptotic behavior of the model. 
 For this, we have to express the cosmological equations governing the evolution of the model as a set of autonomous differential equations. 
 Then the concerned information can be obtained by finding 
 \begin{figure}[h]
\centering
\includegraphics[scale=0.4]{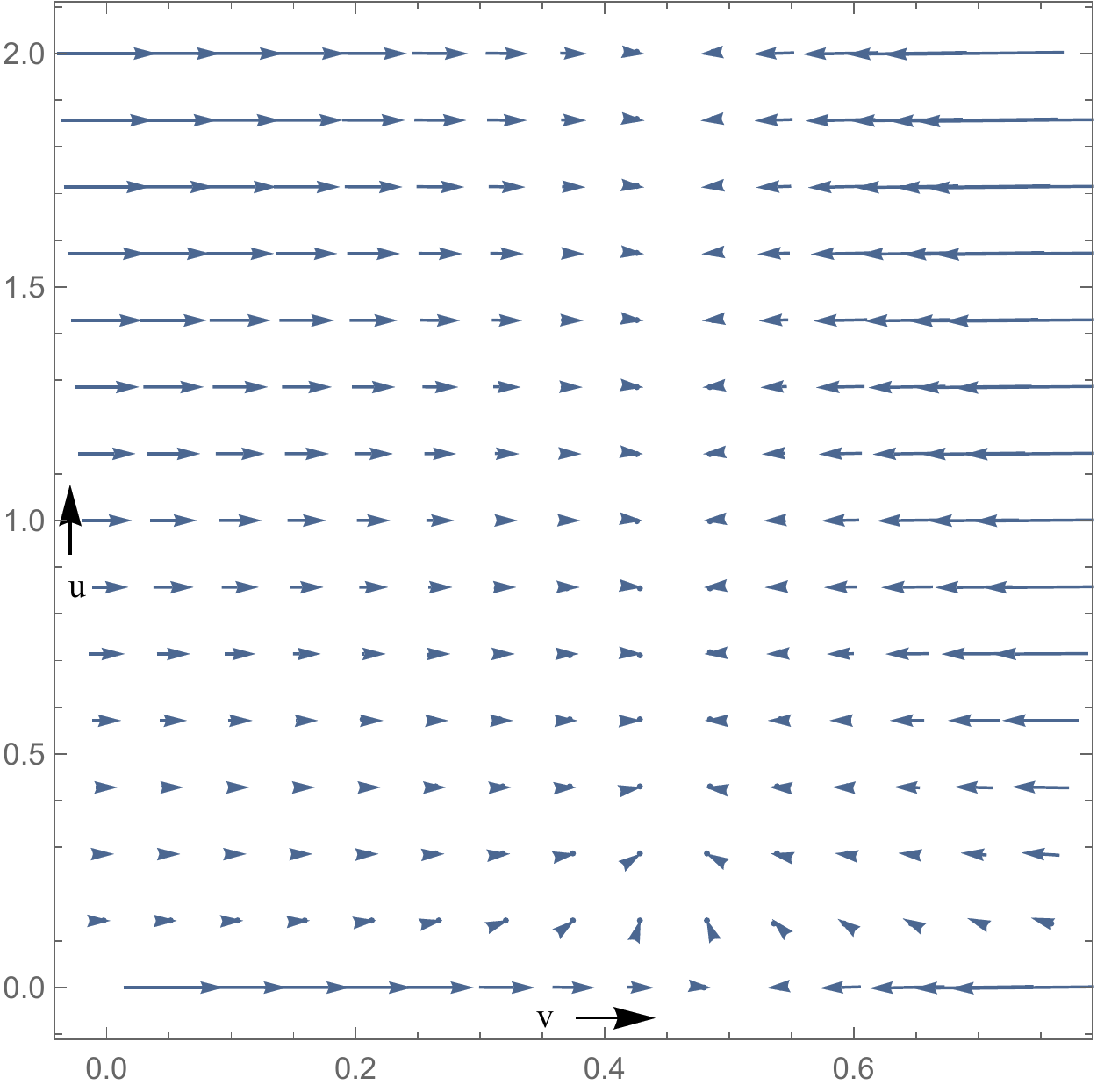}
\centering
\caption{Plot of vector field of the phase space around the critical point(0.607,0.457). The arrowhead of  
trajectories are tilted towards the critical point.} 
\label {fig:PSHR}
\end{figure}
 critical points and analyzing the nature of the trajectories in the neighborhood of the critical points. Eventually it will 
 become clear from this analysis  whether the model in question is consistent with the presently accepted cosmological paradigm. 
 In the present case this means 
 whether the model predicts a stable evolution from a pre-decelerated epoch to a later accelerated epoch.

The first step in the dynamical system analysis is to identify proper phase-space variables. In the present case,  
we define,
\begin{equation} 
u=\Omega_z=(1-\alpha)h^2, \, \,  \, \, \, \,  v=\frac{1}{1+\frac{1}{h}}
\end{equation}
as the phase-space variables where the quantities have their usual meaning. These variables will take the range,
$0\leq u \leq 1$ and $0 \leq v \leq 1.$  The resulting coupled autonomous differential equations can then be formed using the 
Friemdmann equations and they are,
\begin{equation}\label{eqn:2DEquationsOFmotion1}
\dot{u}=6H_0\frac{u}{1-\alpha}\left(\frac{\bar\zeta}{2}-\frac{(1-\alpha)v}{1-v} \right)
\end{equation}
and
\begin{equation}\label{eqn:2DEquationsOFmotion2}
\frac{\dot{v}}{\sqrt{1-\alpha}}=3H_0 v^2\left(\frac{\bar\zeta}{2\sqrt{u}}-\sqrt{1-\alpha}\right).
\end{equation}
\begin{figure}[h]
\centering
\includegraphics[scale=0.4]{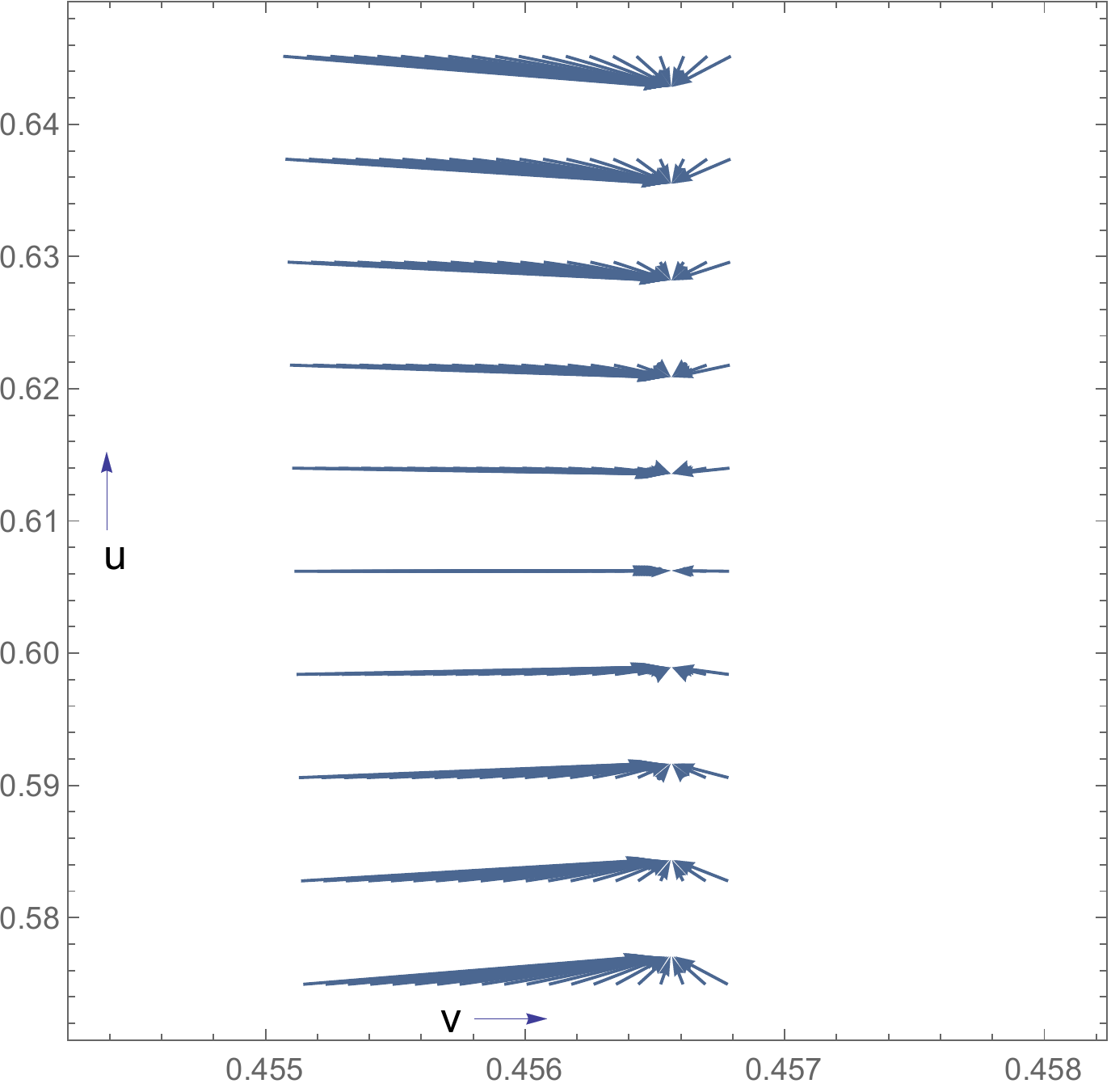}
\caption{Plot of vector field of the phase space close to the critical point.}
\label{fig:vhrVF}
\end{figure}
\begin{figure}[h]
\centering
\includegraphics[scale=0.4]{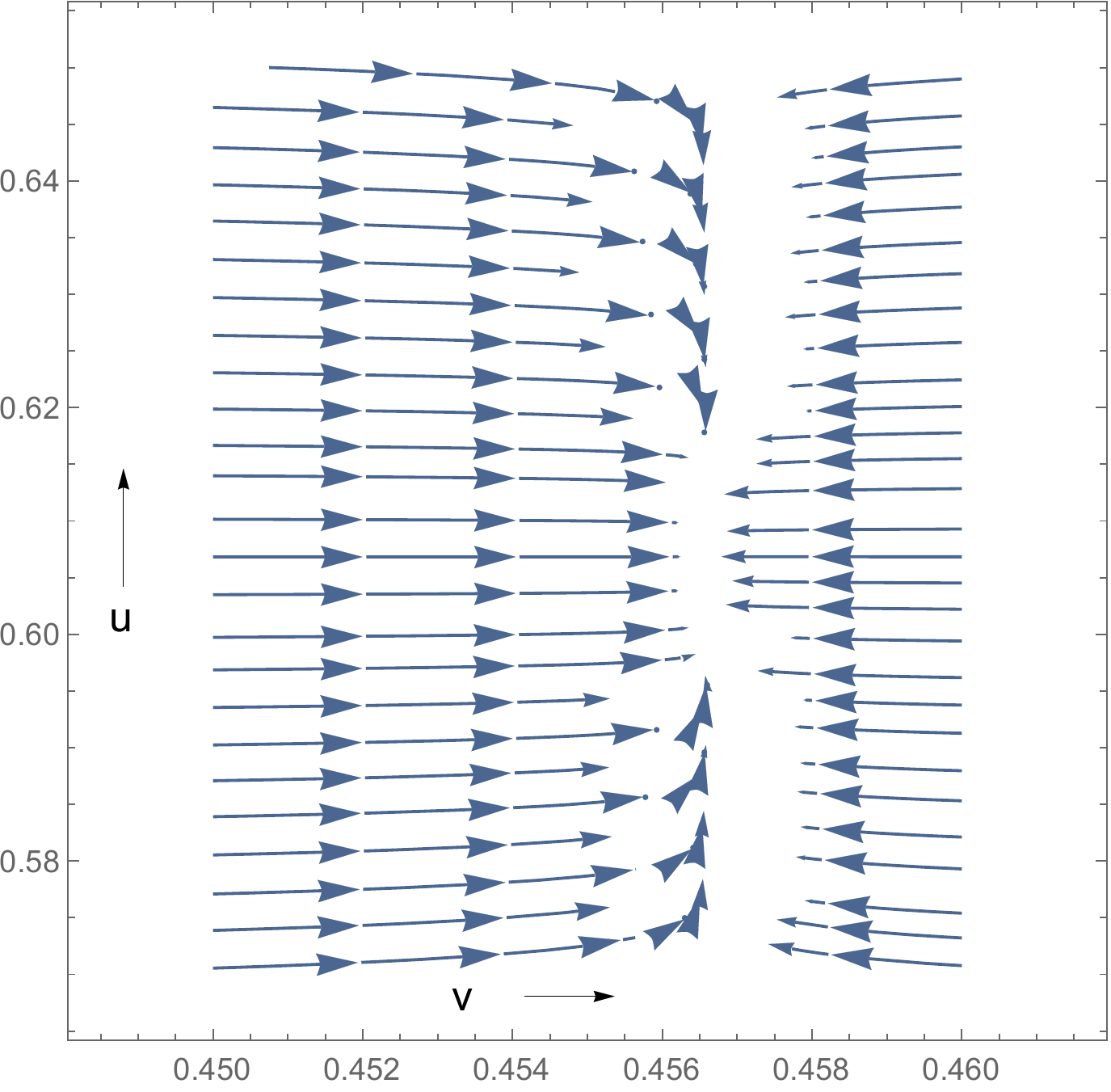}
\caption{Stream plot of the trajectories around the critical point(0.607,0.457)}
\label{fig:vhrSP}
\end{figure} 
The points in the phase space exhibit isomorphism with the exact solutions of cosmological field equations. 
The ODEs are easier to be solved when the derivatives in them are written in terms of $\tau=ln(a).$ The behavior of ODEs in 
the linear, closeby regions of critical points can be expressed in terms of a matrix equation accommodating $\displaystyle u'=\frac{du}{d \tau}$ and 
$\displaystyle v'=\frac{dv}{d \tau}$  enabling flux analysis in terms of $\tau$ parameter of the equivalent autonomous ODE system.  
We then establish the correspondence of the dynamics of $\rho_z=\rho_z(t),\rho_{\Lambda}=\rho_{\Lambda}(t)$ and $H=H(t)$ with the mentioned 
simplified flux analysis in phase space for small perturbations in the linear limit $u\to u+\delta u(\tau)$ and $v\to v+\delta v(\tau),$ where 
$\delta u$ and $\delta v$ are the perturbations.  The critical points are the solutions of the algebraic equations $P(u,v)=0$ and $Q(u,v)=0$ where 
$\displaystyle P(u,v)=6H_0\frac{u}{1-\alpha}\left(\frac{\bar\zeta}{2}-\frac{(1-\alpha)v}{1-v}\right) $ and 
$\displaystyle Q(u,v)=3H_0 v^2\left(\frac{\bar\zeta}{2\sqrt{u}}-\sqrt{1-\alpha}\right).$ 
The perturbations around the critical points satisfy the equation,
\begin{eqnarray}
\left(
\begin{array}{c}
  \delta u^{'} \\ \delta v^{'} 
\end{array}
\right)
= 
\left(
\begin{array}{cc}
\left(\frac{\partial P}{\partial u} \right)_0 & \left(\frac{\partial P}{\partial v} \right)_0  \\
\left( \frac{\partial Q}{\partial u} \right)_0 & \left(\frac{\partial Q}{\partial v}\right)_0
\end{array}
\right)
\left(
\begin{array}{c}
\delta u\\ \delta v
\end{array}
\right)
\end{eqnarray}
where the suffix,$'0'$ denotes the value at the critical point and $2\times2$ matrix in the above equation is the Jacobian. 
The nature of the eigen values of the Jacobian matrix determine the behavior of the system near the critical points.

The critical point of interest corresponding to the equations (\ref{eqn:2DEquationsOFmotion1}) and (\ref{eqn:2DEquationsOFmotion2}) is
$(u_c,v_c)=(0.607,0.457).$ It can be seen that the critical value $v_c$ corresponds to the end de Sitter epoch, where the Hubble parameter 
becomes, $h \to \frac{\bar{\zeta}}{2(1-\alpha)}.$  Then using the relation  $v=\frac{1}{1+\frac{1}{h}}$ and the best fit values of the model parameter, 
it can be
seen that $v_c=0.457$ for the end de Sitter phase. The value $u_c=0.607$, the mass parameter of the bulk viscous Zel'dovich fluid also corresponds 
to the end de Sitter epoch. That is, in the end de Sitter epoch, $u \to \frac{\bar\zeta^2}{4(1-\alpha)}$, which for the 
best estimated values of the model parameters will become equal to 0.607 and is $u_c.$ 
The eigen values corresponding to this critical point is 
 $(0,-4.11).$

The first eigenvalue, $0$ apparently suggests the absence of any isolated
critical point. But it is seen that such a situation does not arise originally from the ODEs by setting $\dot{u}=0$ and $\dot{v}=0.$  
So the apparent discrepancy suggestive of lack of an isolated critical point arises from the errors of linear approximation of trajectory flux 
of the system in its immediate neighborhood. 
A low resolution view of vector field in phase-space is depicted in the figure (\ref{fig:PSHR}). Higher resolution 
vector field plot as in figure (\ref{fig:vhrVF}) makes the view of the critical point as an attractor.
The high resolution stream plot as in figure(\ref{fig:vhrSP}) also is a clear indicator that the critical point is an attractor.

\section{Conclusion}

Many have speculated that matter present in the early stage of the universe were of stiff nature, with equation of state, $p/\rho=1.$ 
But owing to the fast decrease in its 
energy density, it would have effect only on processes like primordial nucleosynthesis. Later works which studied the bulk viscous 
stiff fluid, found that they can even 
cause the late acceleration of the universe. But the main drawback of such models were that, they predicted less age for the universe. 
In the present work we have studied 
a model with Bulk viscous stiff fluid and decaying vacuum energy as cosmic components. We found that the model possesses 
reasonably good back ground evolution, so as 
to produce a late acceleration at about a redshift compatible with the observational results. The model also predicts a
de Sitter epoch as the end phase. It is found that 
the acceleration is mainly due to the effect of bulk viscosity, because for the decaying vacuum to produce a transition 
into the late acceleration, there has to be a constant 
in the vacuum energy density. But the decaying vacuum we have considered doesn't have such a constant. 
The effect of varying  vacuum energy is reflected in the age of the 
universe. It was found that the age of the universe was increased compared to the model with viscous Zel'dovich fluid alone 
as the component. Age obtained is in agreement 
with the age deduced from the observations of the oldest globular clusters.

It was found that the equation of state of the fluid starts form the stiff nature, but eventually reduces to that of the matter and finally 
goes over to that of a 
pure cosmological constant corresponding to the de Sitter epoch. The dynamical system analysis shows that, the end de Sitter phase 
is a stable one. During this stable end 
de Sitter phase, the density of the Zel'dovich fluid is found to be around $0.6,$ which itself confirms that the late 
phase of the universe in this model is dominantly 
controlled by the viscous nature of the Zel'dovich fluid rather than the decaying vacuum energy.

\noindent{Acknowledgement}
One of the authors (KRN) thanks CUSAT Kochi, India for financial support in the form of a research fellowship. We also wish to
thank IUCAA, Pune, India for the local hospitality during our visits, where part of the work had been carried out.






\end{document}